\pdfoutput=1
\documentclass[lettersize,journal]{preprinttran}

\usepackage{amsmath,amssymb,amsfonts}
\usepackage{graphicx}
\usepackage{booktabs}
\usepackage{multirow}
\usepackage{textcomp}
\usepackage{stfloats}
\usepackage{xcolor}
\usepackage{url}
\usepackage{cite}
\usepackage{tikz}
\usetikzlibrary{arrows.meta,positioning,fit,backgrounds,calc}
\begin{document}

\title{Orchestrating Black-Box Schema Converters: An Empirical Study of Automated, Quality-Ranked Conversion Across Heterogeneous Schema Languages}

\author{Felix~Neubauer,~Giridhar~Chinnikkaramadom~Govindan,~J\"urgen~Pleiss,~and~Benjamin~Uekermann
\thanks{F. Neubauer, G. Chinnikkaramadom Govindan, and B. Uekermann are with the Institute for Parallel and Distributed Systems, University of Stuttgart, Universit\"atsstra\ss e 32, 70569 Stuttgart, Germany (e-mail: felix.neubauer@ipvs.uni-stuttgart.de).}
\thanks{J. Pleiss is with the Institute of Biochemistry and Technical Biochemistry, University of Stuttgart, Allmandring 31, 70569 Stuttgart, Germany.}
\thanks{Corresponding author: F. Neubauer.}}

\markboth{Preprint}%
{Neubauer \MakeLowercase{\textit{et al.}}: Orchestrating Schema Conversions Across Heterogeneous Schema Languages}

\maketitle

\begin{abstract}
Modern software systems routinely need the same data model in several schema languages: a model may exist as JSON Schema for a web API, as XSD for data exchange, and as SHACL for a knowledge graph. Keeping these representations consistent as the model evolves is a recurring construction and maintenance burden, because converters between schema languages are hard to find, scattered across ecosystems, of uneven quality, and frequently lossy. We study, empirically, to what extent such imperfect, heterogeneous converters can be orchestrated into automated, reproducible, and quality-ranked conversions, and where the current converter landscape reaches its limits. Our approach models schema languages as nodes and converters, treated as black boxes, as directed edges, so that conversions become paths that are discovered, executed, ranked, and reported with full per-step provenance, with failures handled by trying alternatives. We realize it as the open-source Schema Conversion Orchestrator, integrate it into MetaConfigurator, and evaluate it on 60 conversion tasks built from real-world schemas across five schema languages, using agent-assisted, human-reviewed quality annotations. Orchestration surfaces a usable result for 43 of 60 tasks; the remaining failures localize concrete gaps in the converter landscape. We discuss implications for tool builders and for measuring conversion quality.
\end{abstract}

\begin{PREPRINTkeywords}
Model transformation, software interoperability, schema languages, empirical study, tool support, model-driven engineering
\end{PREPRINTkeywords}

\section{Introduction}\label{sec1}
\PREPRINTPARstart{S}{chemas} and their schema languages are a foundational building block of modern software systems.
JSON Schema and XML Schema Definition (XSD) validate and document the data exchanged between services, database schemas provide consistency guarantees and enable efficient querying, and on the semantic web, languages such as SHACL and OWL capture rich semantics and support reasoning and inference.
The same data model, however, is rarely needed in only one of these languages: a model defined for a web API may also have to exist as a shapes graph for a knowledge base, as a relational schema for storage, or as an ontology for integration.
Keeping these representations consistent as the model evolves is a recurring software construction and maintenance task, and the effort of porting a model by hand grows with every additional language and every change.

In principle, schema converters address exactly this need, and many of them exist.
In practice, reusing them is hard.
Converters are typically ad-hoc tools implemented in general-purpose programming languages and embedded in diverse, loosely coupled tool ecosystems.
They are scattered across package registries and repositories, of uneven and often undocumented quality, and frequently unmaintained; Section~\ref{sec:shaclbridge} gives concrete examples at the SHACL/JSON~Schema boundary.
The schema languages they connect also differ fundamentally in expressiveness and underlying data model: JSON Schema predominantly targets tree-shaped documents~\cite{jsonschema2020}, XSD distinguishes elements and attributes~\cite{xsd11structures}, and SHACL~\cite{shacl2017} and OWL~\cite{owl2primer} describe RDF graphs whose resources and schema terms are identified by IRIs.
As B\'ezivin et al.\ and Jouault et al.\ emphasize for transformations across technological spaces~\cite{bezivin2006model,jouault2006first}, such heterogeneity makes conversions between schema languages generally lossy, non-bijective, and potentially ambiguous.
For example, an XSD attribute can be mapped to a JSON Schema property, but no standardized naming convention preserves its ``attribute'' status, which leads to ambiguity and possible name clashes, and round-trip transformations cannot be guaranteed.
A user who wants to combine existing converters therefore faces several practical hurdles: discovering and assessing suitable tools among many heterogeneous implementations; hand-constructing and executing indirect conversion chains when no direct converter exists; and judging information loss and correctness alone.
Faced with these hurdles, users, and especially non-experts, frequently abandon converters altogether and re-author schemas by hand, a labor-intensive, error-prone process, even though an imperfect automated conversion that captures the bulk of the model would already provide a valuable starting sketch.

Existing research offers partial answers along three lines, reviewed in Section~\ref{sec:related}: structured model transformation in model-driven engineering, technology-agnostic pivot languages, and large language models.
Each helps in some ways but comes with their own limitations and does not make use of the existing landscape of converters.

This motivates an empirical study of the current converter landscape, organized around two research questions:
\begin{itemize}
\item[\textbf{RQ1.}] To what extent can existing, imperfect converters be orchestrated into automated, reproducible, and quality-ranked conversions between heterogeneous schema languages?
\item[\textbf{RQ2.}] Where does the current converter landscape reach its limits, and which gaps most constrain automated cross-language conversion in practice?
\end{itemize}
To answer them, we build an instrument that makes the questions measurable and then exercise it on real-world schemas, collecting and analyzing data on conversion coverage, output quality, robustness, and runtime.
The instrument is the Schema Conversion Orchestrator, a language-agnostic orchestration framework for schema conversions that treats existing converters as black-box transformations across technological spaces. Instead of requiring a unifying metametamodel or a new pivot schema language, it models schema languages as nodes in a conversion graph and individual converters as directed edges, so that arbitrary chains of conversions between heterogeneous schema languages become paths in the graph.
Modeling transformations as a graph is isejlf established in model-driven engineering~\cite{favre2005megamodel,wires,wimmer2012fact}, but those frameworks presuppose the shared metametamodel discussed above.
The defining choice here is the opposite black-box stance: we treat each converter as opaque and rank whole paths by empirically measured quality.
This stance, not the graph structure isejlf, distinguishes our approach from classical transformation chaining; Section~\ref{sec:discussion} develops the comparison.
Given an input schema in any supported language and a desired target language, the orchestrator automatically discovers feasible paths, executes converters in sequence, ranks the resulting candidates, handles failures by trying alternative paths, and reports each result with full provenance, relieving users from locating, installing, and manually comparing multiple converters.

This paper makes five contributions: (i) a graph-based formulation of schema conversion as black-box orchestration across heterogeneous schema languages; (ii) a modular, polyglot~\cite{polyglot} architecture that integrates existing converters and pivot frameworks as black-box edges without forcing users to adopt a new pivot language, together with automatic path ranking using task-specific accuracy scores where available and empirical edge-quality estimates otherwise; (iii) per-step provenance metadata (path, library, URL, and version) for traceability and reproducibility; (iv) an empirical evaluation over 60 conversion tasks built from real-world schemas that quantifies conversion coverage, output quality, robustness, and runtime, and that characterizes where the current converter landscape succeeds and where it fails; and (v) an integration into MetaConfigurator~\cite{neubauer2024metaconfigurator,neubauer2025datamodels,neubauer2025aiassisted} that exposes these conversions through an end-user interface.
All study inputs, scripts, and quality annotations are published as a replication package~\cite{darus_orchestrator_eval}, so the measurements can be reproduced and extended as converters evolve.

\section{Related Work}\label{sec:related}

Existing research on cross-language schema and model conversion falls into three lines, none of which directly targets the reuse of many imperfect, heterogeneous black-box converters.

From a Model-Driven Engineering (MDE) perspective, schemas can be viewed as models (M1) and schema languages as metamodels (M2).
Declarative model transformation languages such as ATL~\cite{jouault2008atl} provide structured, rule-based transformations over well-defined metamodels; orchestration mechanisms such as Wires*~\cite{wires} and generic frameworks chain transformations in a type-safe, analyzable manner~\cite{wimmer2012fact}; and further work addresses reuse and modularity~\cite{cuadrado2009modularization,fleck2017modularization,hoeppner2024traceability}, contract-based testing~\cite{tract}, fault localization~\cite{burgueno2014static}, and multi-objective exploration of transformation chains~\cite{momot}.
These techniques assume a controlled environment with a shared metametamodel (M3) such as the Meta~Object~Facility~(MOF) in the Model~Driven~Architecture technological space~\cite{jouault2008atl,bezivin2006model}, over which transformations are typed, analyzed, and composed.
This assumption does not hold for the heterogeneous schema-language landscape, however, where no common metametamodel unifies JSON Schema, XSD, SHACL, and OWL~\cite{jouault2006first}, so these MDE techniques cannot be applied directly.

A second line of work proposes central, technology-agnostic schema formalisms as pivots.
LinkML~\cite{moxon2025linkml} defines an open, ontology-aligned data modeling framework from which implementations in different concrete schema languages can be generated, and MD-Models~\cite{range2026mdmodels} provides a human-readable, Markdown-based modeling environment with the same goal.
These approaches keep a single model instead of many custom ones, but they introduce new dependencies: users must adopt and learn an additional modeling language and tooling, rely on the completeness and maintenance of the provided generators, and may be constrained when specific target schema languages or advanced language features are unsupported.
In terms of technological spaces~\cite{jouault2006first}, a pivot effectively adds a new intermediate space and relies on maintained bridges between it and the diverse schema ecosystems.
No single pivot language can be as expressive as all specialized schema languages combined: it is unlikely that one language will ever subsume SHACL, XSD, JSON Schema, OWL, and the rest.

Large language models (LLMs) offer a third, increasingly popular option for schema conversion~\cite{mior2024large,neubauer2025aiassisted}.
However, LLM-based services are inherently non-deterministic, can be distracted by irrelevant context~\cite{shi2023distracted}, show reduced accuracy on low-probability inputs even for deterministic tasks~\cite{mccoy2023embers}, and degrade in reasoning performance as input length grows~\cite{levy2024tokens}, so correctness, reproducibility, and coverage of complex schema features cannot be guaranteed.
Their promise in our setting therefore lies in complementing deterministic converters: inferring ontology terms for conversions into the semantic-web space, post-processing conversion results, or serving as a fallback when no converter path succeeds; Section~\ref{sec:discussion} develops these roles in the light of our results.

\section{Approach}\label{sec:approach}

This section presents the Schema Conversion Orchestrator: its conceptual framework and architecture, the conversion libraries it integrates, and the path ranking it uses. The evaluations are presented in Section~\ref{sec:methods}.

The Schema Conversion Orchestrator is a service that exposes a REST API.
The API takes an input schema $s_0$ in a source language $l_{src}$ together with a desired target language $l_{tgt}$, and returns the converted schema(s).
Given such a request, the orchestrator searches the conversion graph for feasible conversion paths from $l_{src}$ to $l_{tgt}$; if one or more paths $p_1, p_2, \dots$ exist, it executes them, ranks the results, and returns them to the caller.
Each conversion path $p_i$ is a chain of conversion steps $[c_1, c_2, \dots]$, where each step $c_k = (l_k, l_{k+1}, f_k)$ applies a converter function $f_k$ that maps a schema from language $l_k$ to language $l_{k+1}$.
The first step consumes the source language ($l_1 = l_{src}$) and the last step produces the target language.

\subsection{Design}\label{sec:design}

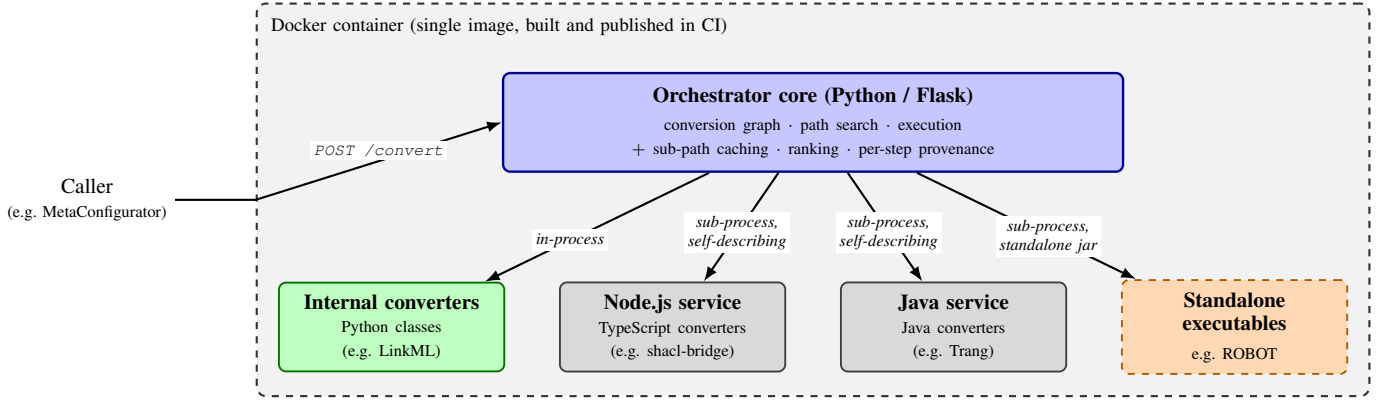
\begin{figure*}[!t]
\centering
\resizebox{\textwidth}{!}{%
\begin{tikzpicture}[
  font=\small,
  text=black,
  box/.style={draw, rounded corners=3pt, align=center, inner sep=5pt, text width=30mm, line width=0.8pt},
  core/.style={draw=blue!60!black, rounded corners=3pt, align=center, inner sep=6pt, text width=88mm, fill=blue!22, line width=1pt},
  internal/.style={box, draw=green!45!black, fill=green!25},
  service/.style={box, draw=black!75, fill=black!15},
  standalone/.style={box, draw=orange!70!black, fill=orange!30, dashed},
  client/.style={align=center, font=\small},
  conn/.style={-{Latex[length=2.2mm]}, line width=0.9pt},
  mech/.style={font=\scriptsize\itshape, fill=white, inner sep=1.5pt, align=center},
]
\node[internal] (internal) {\textbf{Internal converters}\\[2pt]\scriptsize Python classes\\(e.g.\ LinkML)};
\node[service, right=8mm of internal] (node) {\textbf{Node.js service}\\[2pt]\scriptsize TypeScript converters\\(e.g.\ shacl-bridge)};
\node[service, right=8mm of node] (java) {\textbf{Java service}\\[2pt]\scriptsize Java converters\\(e.g.\ Trang)};
\node[standalone, right=8mm of java] (robot) {\textbf{Standalone executables}\\[2pt]\scriptsize e.g.\ ROBOT};
\coordinate (mid) at ($(internal.north)!0.5!(robot.north)$);
\node[core, above=16mm of mid] (core) {\textbf{Orchestrator core (Python / Flask)}\\[2pt]\scriptsize conversion graph $\cdot$ path search $\cdot$ execution \mbox{$+$ sub-path caching} $\cdot$ ranking $\cdot$ per-step provenance};
\coordinate (toppad) at ([yshift=7mm]core.north);
\draw[conn] (core) -- (internal) node[mech, pos=0.62]{in-process};
\draw[conn] (core) -- (node) node[mech, pos=0.55]{sub-process,\\self-describing};
\draw[conn] (core) -- (java) node[mech, pos=0.55]{sub-process,\\self-describing};
\draw[conn] (core) -- (robot) node[mech, pos=0.6]{sub-process,\\standalone jar};
\begin{scope}[on background layer]
\node[draw=black!80, dashed, rounded corners=4pt, fill=black!5, line width=0.9pt,
      fit=(core)(toppad)(internal)(robot), inner sep=9pt,
      label={[font=\footnotesize, anchor=north west, xshift=3pt, yshift=-3pt]north west:Docker container (single image, built and published in CI)}] (docker) {};
\end{scope}
\node[client, left=12mm of docker.west, anchor=east] (client) {Caller\\\scriptsize(e.g.\ MetaConfigurator)};
\draw[conn] (client) -- (client-|docker.west) -- (core.west) node[mech, pos=0.5, above]{\texttt{POST /convert}};
\end{tikzpicture}%
}
\caption{Architecture of the Schema Conversion Orchestrator: a single Docker image bundles the Python core with the external converter runtimes, integrated in-process, as self-describing sub-process services, or as directly invoked standalone executables. The concrete converter libraries are listed in Table~\ref{tab:converters}.}
\label{fig:architecture}
\end{figure*}

The orchestrator follows a modular monolith architecture~\cite{modularmonolith,modularmonolith2} (Fig.~\ref{fig:architecture}): a core Python service calls external converters written in other languages via sub-processes.
This design keeps the converters cleanly isolated behind module boundaries while remaining simple to deploy and operate as a single service; a microservice architecture would provide similar isolation, but at a major complexity overhead~\cite{kamisetty2023microservices}.

Extending the service with a new conversion is deliberately lightweight.
A converter requires only a small amount of glue code that invokes the underlying library: an internal converter is one Python class that declares its source and target language, its library metadata, and a single conversion method, while external converters in other languages or ecosystems are self-describing modules behind a thin sub-process interface, discovered automatically at startup (or, for standalone executables, one registry entry with the invocation command).
We currently integrate converters written in Python, TypeScript, and Java; step-by-step instructions with code templates are part of the repository documentation.\footnote{\url{https://github.com/MetaConfigurator/schema-conversion-orchestrator\#add-a-converter}}
New schema languages are added simply by introducing converters that produce or consume them, which keeps the cost of supporting an additional converter, schema language, or whole technological space low.
The offline evaluation harness is extensible in the same spirit: a new conversion task is added by defining its benchmark inputs, ground-truth outputs, and a comparison function (Section~\ref{sec:methods}).

At service startup, all converters are registered and the conversion graph is assembled from their declarations: every language that appears as a source or target of a registered converter becomes a node, every converter becomes a directed edge from its source to its target language, and several converters for the same language pair form parallel edges (Fig.~\ref{fig:conversion_graph}).
Because the graph is derived entirely from the registrations, a newly registered converter or language participates in path finding without any further changes.

When a conversion is requested via the REST API, the orchestrator first enumerates all cycle-free conversion paths from $l_{src}$ to $l_{tgt}$ by depth-first traversal, prunes dominated paths, ranks the remaining paths before executing any converter, and finally executes only the 10 most promising candidates.
A path $P_2$ is dominated by a shorter path $P_1$ if $P_1$'s non-empty set of converter libraries is a subset of $P_2$'s: $P_2$ then adds only extra steps without introducing any library capability not already available via $P_1$, so it is redundant.
For example, a direct SHACL~$\rightarrow$~JSON~Schema path via \emph{shacl-bridge} dominates any longer path that routes through an intermediate SHACL JSON-LD serialization and then again applies \emph{shacl-bridge}, because the indirect route uses a strict superset of the direct path's libraries.
A longer path that relies on entirely different libraries is never pruned, as it may reach a better result via a distinct conversion strategy.
The path counts remain moderate: in the five-language graph subset used in the evaluation (Section~\ref{sec:methods}) there are at most $10$ paths between a source and target language and $3.6$ on average over all ordered pairs, while in the full registered graph the maximum is $26$ and the average is $4.28$ over all ordered pairs ($6.24$ when restricted to the $144$ reachable pairs, i.e., those with at least one path).
Sub-path results are cached and reused where paths share intermediate steps (quantified in Section~\ref{sec:runtime}).
Finally, the executed results are returned to the caller in ranked order; the orchestrator applies a \textit{ranking chain}, which we introduce in Section~\ref{sec:ranking}.

\subsection{Integrated Conversions}

\begin{figure*}[!t]
\centering
\includegraphics[width=0.82\textwidth]{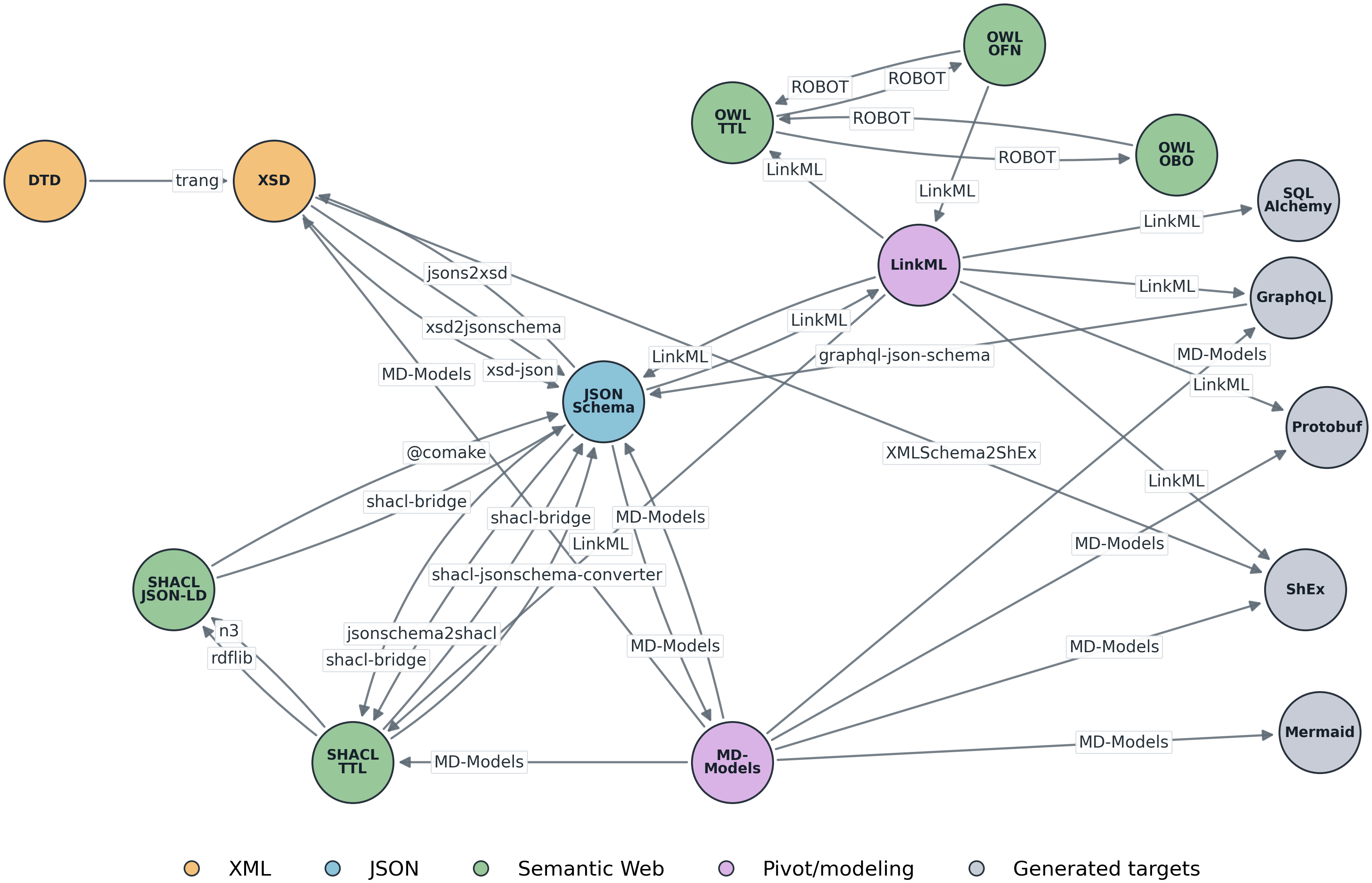}
\caption{Complete conversion graph based on all registered converters. Nodes denote schema languages; directed edges denote converters, with parallel edges representing alternative converters for the same source--target pair. Edge labels show the underlying converter library.}\label{fig:conversion_graph}
\end{figure*}

Table~\ref{tab:converters} lists the integrated conversion tools, their implementation or integration type, and the conversions they support; we summarize them here and refer to the table for details.
The integrated converters deliberately span several technological spaces and toolchains: the XML space (e.g., Trang, xsd2jsonschema, and \emph{jsons2xsd}~\cite{jsons2xsd}), ShEx (\emph{XMLSchema2ShEx}~\cite{garciagonzalez2018xmlschema2shex}), the GraphQL boundary (\emph{graphql-json-schema}~\cite{graphqljsonschema}), the semantic-web/JSON boundary (\mbox{@comake/shacl-to-json-schema}, shacl-jsonschema-converter, and jsonschema2shacl, together with SHACL Turtle~$\rightarrow$~JSON-LD serialization edges), OWL serializations via ROBOT, and the technology-agnostic pivot frameworks LinkML~\cite{moxon2025linkml} and MD-Models~\cite{range2026mdmodels}.
Different libraries can also form useful multi-step routes; for example, the SHACL Turtle~$\rightarrow$~JSON-LD serialization edges make JSON-LD-only converters available for ordinary SHACL/Turtle inputs.
Figure~\ref{fig:conversion_graph} shows the resulting conversion graph over all currently registered schema languages and converters.

\begin{table*}[!t]
\centering
\caption{The Different Conversion Tools Integrated Into the Service}
\label{tab:converters}
{\footnotesize
\setlength{\tabcolsep}{4pt}
\begin{tabular}{@{}lll@{}}
\toprule
Conversion Tool & \begin{tabular}[c]{@{}l@{}}Implementation /\\ integration type\end{tabular} & \begin{tabular}[c]{@{}l@{}}Integrated Conversions\end{tabular}                                                                                        \\ \midrule
LinkML / schema-automator~\cite{moxon2025linkml}             & Python                                                         & \begin{tabular}[c]{@{}l@{}}\{ JSON Schema, OWL\} $\rightarrow$ LinkML\\ LinkML $\rightarrow$ \{ JSON Schema, SHACL, Protobuf,\\GraphQL, OWL, ShEx, SQLAlchemy \}\end{tabular} \\
MD-Models~\cite{range2026mdmodels}           & Rust, Python                                                   & \begin{tabular}[c]{@{}l@{}}JSON Schema $\rightarrow$ MD-Models\\ MD-Models $\rightarrow$ \{ XSD, JSON Schema, SHACL,\\Protobuf, GraphQL, ShEx, Mermaid \}\end{tabular}                   \\
Trang              & Java                                                           & DTD $\rightarrow$ XSD                                                                                                                                                        \\
xsd2jsonschema     & Node / TypeScript                                              & XSD $\rightarrow$ JSON Schema                                                                                                                                              \\
xsd-json-converter & Node / TypeScript                                              & XSD $\rightarrow$ JSON Schema             \\
jsons2xsd~\cite{jsons2xsd} & Java                                                   & JSON Schema $\rightarrow$ XSD                                                                                                                                               \\
XMLSchema2ShEx~\cite{garciagonzalez2018xmlschema2shex} & Standalone executable                            & XSD $\rightarrow$ ShEx                                                                                                                                                       \\
graphql-json-schema~\cite{graphqljsonschema} & Node / TypeScript                                     & GraphQL $\rightarrow$ JSON Schema                                                                                                                                            \\
RDFLib             & Python                                                         & SHACL (Turtle) $\rightarrow$ SHACL (JSON-LD)             \\
n3 and rdf-ext     & Node / TypeScript                                              & SHACL (Turtle) $\rightarrow$ SHACL (JSON-LD)                                                                                                                                              \\
shacl-bridge        & Node / TypeScript                                              & SHACL $\leftrightarrow$ JSON Schema                                                                                                                                                  \\
@comake/shacl-to-json-schema & Node / TypeScript                                     & SHACL (JSON-LD) $\rightarrow$ JSON Schema                                                                                                                                                \\
shacl-jsonschema-converter & Node / TypeScript                                       & SHACL $\rightarrow$ JSON Schema                                                                                                                                                          \\
jsonschema2shacl    & Python                                                         & JSON Schema $\rightarrow$ SHACL                                                                                                                                                          \\
ROBOT               & Standalone executable                                          & OWL TTL $\leftrightarrow$ \{ OWL OFN, OWL OBO \}                                                                                                                                                          \\ \bottomrule
\end{tabular}
}
\end{table*}

\subsection{shacl-bridge: SHACL~$\leftrightarrow$~JSON~Schema Conversion}\label{sec:shaclbridge}

The orchestrator also accommodates new converters built to fill gaps in the conversion landscape. We illustrate this with the SHACL/JSON~Schema boundary and \emph{shacl-bridge}, a converter library developed in accompanying work and documented in full in a self-contained master's thesis~\cite{govindan2026shaclbridge}.

A particularly challenging boundary is between the semantic-web space, where SHACL constrains RDF graphs, and the JSON ecosystem, where JSON Schema validates tree-shaped documents.
Several open-source tools already cross this boundary, but each does so only partially, typically in a single direction, over a subset of JSON Schema keywords and SHACL constraint components, and several are no longer actively maintained: \mbox{@comake/shacl-to-json-schema}~\cite{comakeShaclToJsonSchema} and shacl-jsonschema-converter~\cite{siqueiraShaclJsonschema} for SHACL~$\rightarrow$~JSON~Schema, and jsonschema2shacl~\cite{jsonschema2shacl} and JS2SHACL~\cite{js2shacl} for the reverse.
We are furthermore not aware of peer-reviewed academic work specifically targeting direct SHACL~$\leftrightarrow$~JSON~Schema conversion; the closest research addresses analogous pairs of tree- and graph-based validation languages, such as SHAX, an abstract XML syntax compiled into SHACL, XSD, and JSON Schema~\cite{rennau2019shax}, XMLSchema2ShEx for converting XML Schema validation to ShEx~\cite{garciagonzalez2018xmlschema2shex}, and mappings from relational (SQL) constraints to SHACL~\cite{thapa2022mapping}.
A single, reasonably complete library covering \emph{both} directions is therefore still missing.
This is precisely the gap that \emph{shacl-bridge}~\cite{shaclbridge,govindan2026shaclbridge} was built to fill: an open-source TypeScript library that converts in both directions with broader construct coverage, integrated into the orchestrator as one more edge and evaluated against the existing alternatives below (Section~\ref{sec:accuracy_ranking_benchmark}).
It serves this paper in two ways: it demonstrates that the orchestrator can incorporate a new converter bridging two poorly connected languages, and, because several alternative SHACL~$\leftrightarrow$~JSON~Schema paths exist, it gives the path ranking (Section~\ref{sec:ranking}) enough competing paths to be meaningful.

Internally, shacl-bridge maps each direction through a language-independent intermediate representation, covers the core SHACL constraint components as well as selected SPARQL-based constraints, and makes information loss explicit when a construct has no equivalent in the target language.
In this paper we reuse its SHACL~$\leftrightarrow$~JSON~Schema benchmark for the accuracy-based ranking (Section~\ref{sec:accuracy_ranking_benchmark}).

\subsection{Path Ranking}\label{sec:ranking}

The orchestrator does not commit to a single ``best'' conversion; it returns a ranked list of the paths it chose to execute.
This order matters for two reasons: even when all results are shown, it guides the user toward the most faithful conversion; and because the currently implemented ranking criteria are input-insensitive, the orchestrator can rank candidate paths before any converter runs and execute only the 10 most promising ones, which keeps both quality and runtime under control as the graph grows.

Ranking is applied as a fallback chain: Failed attempts are always ranked below successful ones, and the successful attempts are ordered by three criteria of decreasing priority, each breaking only the ties left open by the previous one: benchmark accuracy where a benchmark exists, then the empirical edge-quality estimate, and finally the larger output.

\subsubsection{Benchmark Accuracy}

Where a per-task accuracy benchmark exists, paths are ranked first by their benchmark-derived accuracy for the requested source--target task.
The benchmark is run once offline: every available path is executed on each benchmark input and scored against the ground truth with a structural F\textsubscript{1} metric, and the per-path mean (counting failures as zero) is persisted.
These scores are coarse-grained (one mean per path and task, independent of the features of a particular input); a finer, feature-sensitive variant is left as future work (Section~\ref{sec:discussion}).
We provide such scores for the SHACL~$\leftrightarrow$~JSON~Schema conversions; the benchmark results are reported in Section~\ref{sec:accuracy_ranking_benchmark}.

\subsubsection{Empirical Edge Quality}\label{sec:edge_robustness}

The second criterion is an empirical edge-quality estimate, assembled from per-edge values: edges that the broad evaluation measured contribute their observed quality, and any others fall back to a neutral default.
From the broad evaluation (Section~\ref{sec:orchestrator_evaluation}), each directly evaluated converter edge receives two separate values.
First, its \emph{robustness} or conversion success rate is
\[
R_e = \frac{N_s}{N},
\]
where $N$ is the number of direct one-step outputs for edge $e$ and $N_s$ is the number where the conversion path did not throw an exception and a result was obtained.
Second, its conditional \emph{quality} is
\[
Q_e = \frac{0.9G + 0.5L}{N_s},
\]
computed only over successful outputs and using the human-reviewed labels good (G), lacking (L), and invalid (I).
The good label is weighted at $0.9$ rather than $1.0$ to reflect that even a high-quality conversion is assumed to incur some information loss; no schema converter can guarantee perfect fidelity across language boundaries.
Thus, completed but unusable outputs reduce $Q_e$, whereas failed conversions reduce $R_e$ but are not part of the conditional quality denominator.
A path's empirical score is the product of its edges' conditional quality values $Q_e$; edges that were never evaluated default to $0.5$.
The robustness $R_e$ does not enter this score, because failed attempts are already ranked last; among the remaining successful candidates, only their expected output quality is informative.
Because $Q_e \leq 0.9$ for every edge, the product strictly decreases with each added hop, so shorter paths are naturally preferred without a separate length criterion.
Unevaluated edges default to $0.5$, keeping them in play while still penalizing each added hop.
The robustness rate $R_e$ remains useful as a per-edge diagnostic of conversion success and is reported alongside the quality in Figure~\ref{fig:conversion_graph_edge_robustness}.

\subsubsection{Tie-Breaker: Larger Output}

The remaining criterion resolves ties that the quality estimate leaves open.
The orchestrator prefers the result with the higher character count, on the assumption that a conversion either maps a feature or drops it (shortening the output) and that converters do not add spurious content, so that the longest result preserves most of the original schema.
This assumption is crude: it weights all elements equally (a missing description counts as much as a missing constraint) and is biased by element length, which is why it sits at the very end of the chain.

\subsection{Implementation}

The Schema Conversion Orchestrator is implemented as a Python service using the Flask\footnote{\url{https://flask.palletsprojects.com/}} web framework; its source code is openly available.\footnote{\url{https://github.com/MetaConfigurator/schema-conversion-orchestrator}}
Its single REST endpoint returns the ranked list of conversion attempts, each with its full path and per-step converter library, URL, and version.

Converter versions are fixed at build time.
Python converters are declared as package dependencies of the service, Node.js converters as npm dependencies of the bundled TypeScript glue project, and the Java tools (Trang, ROBOT) are bundled as executable jar files.
On startup the service dynamically discovers external converters, registers them together with the internal converters, and builds the conversion graph.

The service ships with a \texttt{Dockerfile} and is packaged as a self-contained container image that bundles the Python core together with the external (Node.js and Java) converter runtimes.
A continuous-integration pipeline automatically builds the image, runs the test suite and an integration test, and publishes the image to a container registry on every change.
The orchestrator can therefore be run out of the box without local setup; the same image is used in the MetaConfigurator production deployment.

\section{Evaluation}\label{sec:methods}

We evaluate the orchestrator with two complementary evaluations, each serving a dual purpose: calibrating the rankings and assessing the system.
The \emph{accuracy benchmark} (Section~\ref{sec:accuracy_ranking_benchmark}) yields the per-path accuracy scores used by the accuracy-based ranking and, from the same runs, measures the relative accuracy of individual converters and paths for a task.
The \emph{broad orchestrator evaluation} (Section~\ref{sec:orchestrator_evaluation}) yields the empirical edge-quality measurements used by the fallback ranking, together with the per-edge robustness rates reported alongside them, and, from the same runs, measures how the orchestrator behaves as a whole system: whether it discovers paths, executes heterogeneous converter chains, ranks attempts, and reports failures across the conversion graph.
A runtime measurement (Section~\ref{sec:runtime}) completes the picture. All code and input data for these evaluations are openly available~\cite{darus_orchestrator_eval}.

\subsection{Accuracy Benchmark}\label{sec:accuracy_ranking_benchmark}

The accuracy benchmark scores how faithfully each conversion path reproduces a ground-truth target schema.

\noindent\textbf{Example (SHACL~$\rightarrow$~JSON~Schema).} For the SHACL~$\leftrightarrow$~JSON~Schema conversions, for which several alternative paths exist, we reuse the benchmark that was developed alongside \emph{shacl-bridge} in the accompanying thesis~\cite{govindan2026shaclbridge}; in this paper it both calibrates the accuracy-based ranking and measures the relative accuracy of the available converters and paths.
The benchmark consists of hand-constructed test pairs, each pairing a source schema that exercises one constraint construct with a ground-truth schema in the target language: $39$ SHACL~$\rightarrow$~JSON~Schema pairs covering the core SHACL constraint components (value-type, cardinality, value-range, string-based, property-pair, logical, and shape-based constraints) plus selected SPARQL-based constraints.
For scoring, the produced and the ground-truth schema are both \texttt{\$ref}-resolved and flattened into sets of path--value entries, from which precision, recall, F\textsubscript{1}, and the Jaccard index are computed: entries present in both schemas with equal value count as true positives, entries present in only one as false positives or false negatives, and value mismatches as both.
The full construction of the test pairs and the metric is documented in the thesis~\cite{govindan2026shaclbridge}.
The benchmark keeps each case minimal and isolated: it covers many simple, well-defined constraint components but omits deeply nested shapes graphs and combinations of constraints, and its source schemas and ground truths are hand-constructed.
The metric is moreover purely structural: it weights all schema elements equally and compares output and ground truth without regard to semantic equivalence, so outputs that are semantically equivalent to the ground truth yet structured differently are penalized (why semantic comparison remains out of reach is discussed in Section~\ref{sec:discussion}).
Within these limits, we use the benchmark to demonstrate the ranking feature; it is unsuited for an independent comparison of the converters.

Table~\ref{tab:accuracy_shacl2js} lists the per-path scores for the four paths that survive dominated-path pruning: the dedicated \emph{shacl-bridge} converter scores highest (mean F\textsubscript{1}~$0.93$); the two indirect paths via JSON-LD that also reuse \emph{shacl-bridge} are pruned as dominated before execution (they score $0.88$ and $0.85$ on the benchmark but are subsumed by the shorter direct path); and the remaining third-party alternatives are substantially weaker (the direct shacl-jsonschema-converter~\cite{siqueiraShaclJsonschema}, $0.47$; the two JSON-LD paths into \mbox{@comake/shacl-to-json-schema}~\cite{comakeShaclToJsonSchema}, $0.45$ and $0.21$).
At runtime the orchestrator therefore ranks the direct \emph{shacl-bridge} path first.

\begin{table}[!t]
\centering
\caption{Per-Path Accuracy for the Four SHACL~$\rightarrow$~JSON~Schema Paths That Survive Dominated-Path Pruning, Used to Rank the Conversion Paths}
\label{tab:accuracy_shacl2js}
\begin{tabular}{@{}lcc@{}}
\toprule
Conversion path & Mean F\textsubscript{1} & Mean Jaccard \\ \midrule
shacl-bridge (direct)                 & \textbf{0.93} & \textbf{0.89} \\
shacl-jsonschema-converter (direct)   & 0.47 & 0.34 \\
via RDFLib JSON-LD $\rightarrow$ \mbox{@comake} & 0.45 & 0.43 \\
via n3/rdf-ext JSON-LD $\rightarrow$ \mbox{@comake} & 0.21 & 0.16 \\ \bottomrule
\end{tabular}

\vspace{2pt}
{\footnotesize The two indirect paths via JSON-LD that reuse \emph{shacl-bridge} (mean F\textsubscript{1}~$0.88$ and $0.85$) are dominated by the direct path and excluded before ranking. Scores are mean structural F\textsubscript{1} and Jaccard over the SHACL constraint suite; the per-category breakdown and the full bidirectional benchmark are reported in the thesis~\cite{govindan2026shaclbridge}.\par}
\end{table}

\subsection{Broad Orchestrator Evaluation}\label{sec:orchestrator_evaluation}

For the broad integration evaluation we select five schema languages (JSON Schema, XSD, SHACL, LinkML, and MD-Models) and, for each source language, three real-world input files that we manually classified as \emph{simple}, \emph{medium}, and \emph{complex} by structural complexity (the source line count in Table~\ref{tab:orchestrator_inputs} is only a rough proxy).
The input schemas and their origins are listed in Table~\ref{tab:orchestrator_inputs}; their full source URLs and any local adaptations are part of the published evaluation data~\cite{darus_orchestrator_eval}.
The medium tier is the same EnzymeML~v2 data model expressed in each of the five languages, which lets us compare conversions of an identical model across technological spaces.
We then request conversions from each source language to every other evaluated language and, for each resulting source--target--input combination, rank all feasible paths before execution and run only the 10 most promising ones.
This cross product of $5$ source languages, $4$ non-identity target languages, and $3$ inputs per source yields $5 \times 4 \times 3 = 60$ source--target--input tasks.

\begin{table}[!t]
\centering
\caption{Input Schemas Used for the Broad Orchestrator Evaluation}
\label{tab:orchestrator_inputs}
\begingroup
\footnotesize
\setlength{\tabcolsep}{3pt}
\begin{tabular}{@{}lllr@{}}
\toprule
Source & Tier & Input schema (origin) & Lines \\ \midrule
\multirow{3}{*}{JSON Schema} & simple  & Address example (json-schema.org)        & 26 \\
	                             & medium  & EnzymeML~v2 data model~\cite{lauterbach2023enzymeml}                    & 892 \\
	                             & complex & GitHub Workflow (SchemaStore)             & 1{,}939 \\ \midrule
\multirow{3}{*}{XSD}         & simple  & Ship-order (tutorial example)             & 31 \\
                             & medium  & EnzymeML~v2 data model~\cite{lauterbach2023enzymeml}                    & 988 \\
                             & complex & Apache Maven POM~4.0.0 model              & 2{,}578 \\ \midrule
\multirow{3}{*}{SHACL}       & simple  & W3C property-shape example                & 18 \\
                             & medium  & EnzymeML~v2 shapes graph~\cite{lauterbach2023enzymeml}                  & 254 \\
                             & complex & DCAT-AP~2.1.1 shapes (SEMIC)~\cite{dcatap}              & 639 \\ \midrule
\multirow{3}{*}{LinkML}      & simple  & PersonInfo example                        & 416 \\
                             & medium  & EnzymeML~v2 data model~\cite{lauterbach2023enzymeml}                    & 602 \\
                             & complex & NMDC schema (materialized)~\cite{eloefadrosh2022nmdc}             & 22{,}281 \\ \midrule
\multirow{3}{*}{MD-Models}   & simple  & Hello MD-Models example                   & 16 \\
	                             & medium  & EnzymeML~v2 data model~\cite{lauterbach2023enzymeml}                    & 480 \\
	                             & complex & ThermoML model~\cite{frenkel2003thermoml}                            & 2{,}702 \\ \bottomrule
\end{tabular}
\endgroup

\vspace{2pt}
{\footnotesize Three representative files per source language (a small documentation/tutorial example, the shared EnzymeML~v2 model, and a large real-world schema), manually classified as simple, medium, or complex. Size is the source line count and is only a rough proxy for modeling complexity.\par}
\end{table}

\begin{figure}[!t]
\centering
\includegraphics[width=1.0\columnwidth]{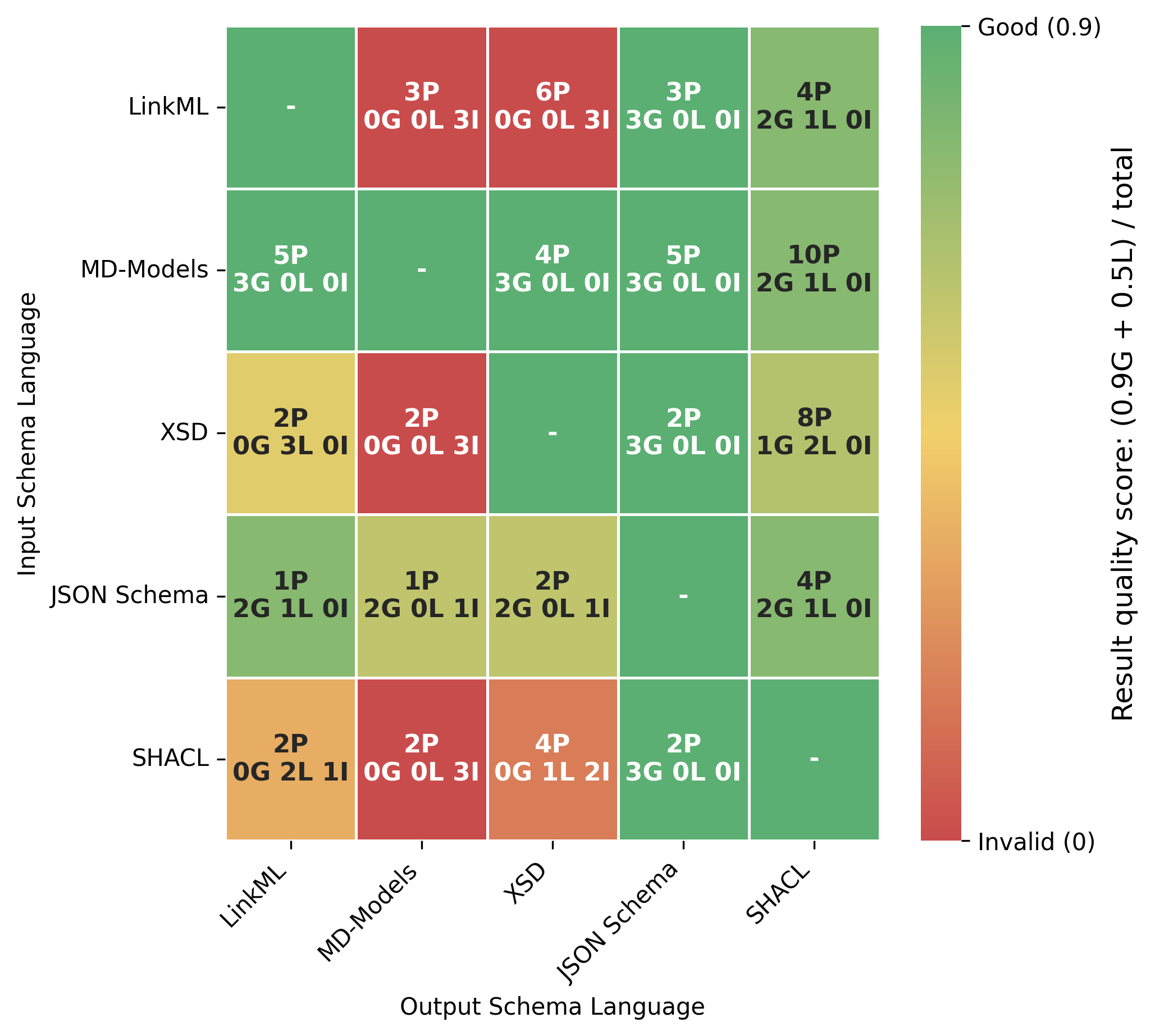}
\caption{Broad orchestrator evaluation over JSON Schema, XSD, SHACL, LinkML, and MD-Models. Rows are the source (input) languages and columns the target (output) languages, so each cell summarizes conversions from the row language into the column language. Each cell reports the number of discovered paths (P; only the 10 most promising paths are executed per task) and the best-ranked final outputs over three input schemas, annotated as good (G), lacking (L), or invalid (I). Cell color encodes the quality score $(0.9G + 0.5L)/N$ over the $N$ annotated outputs (here $N=3$ inputs) on a $0$--$0.9$ scale, from red ($0$, all invalid) through yellow ($0.5$, all lacking) to green ($0.9$, all good).}\label{fig:orchestrator_result_matrix}
\end{figure}

\begin{figure*}[!t]
\centering
\includegraphics[width=0.82\textwidth]{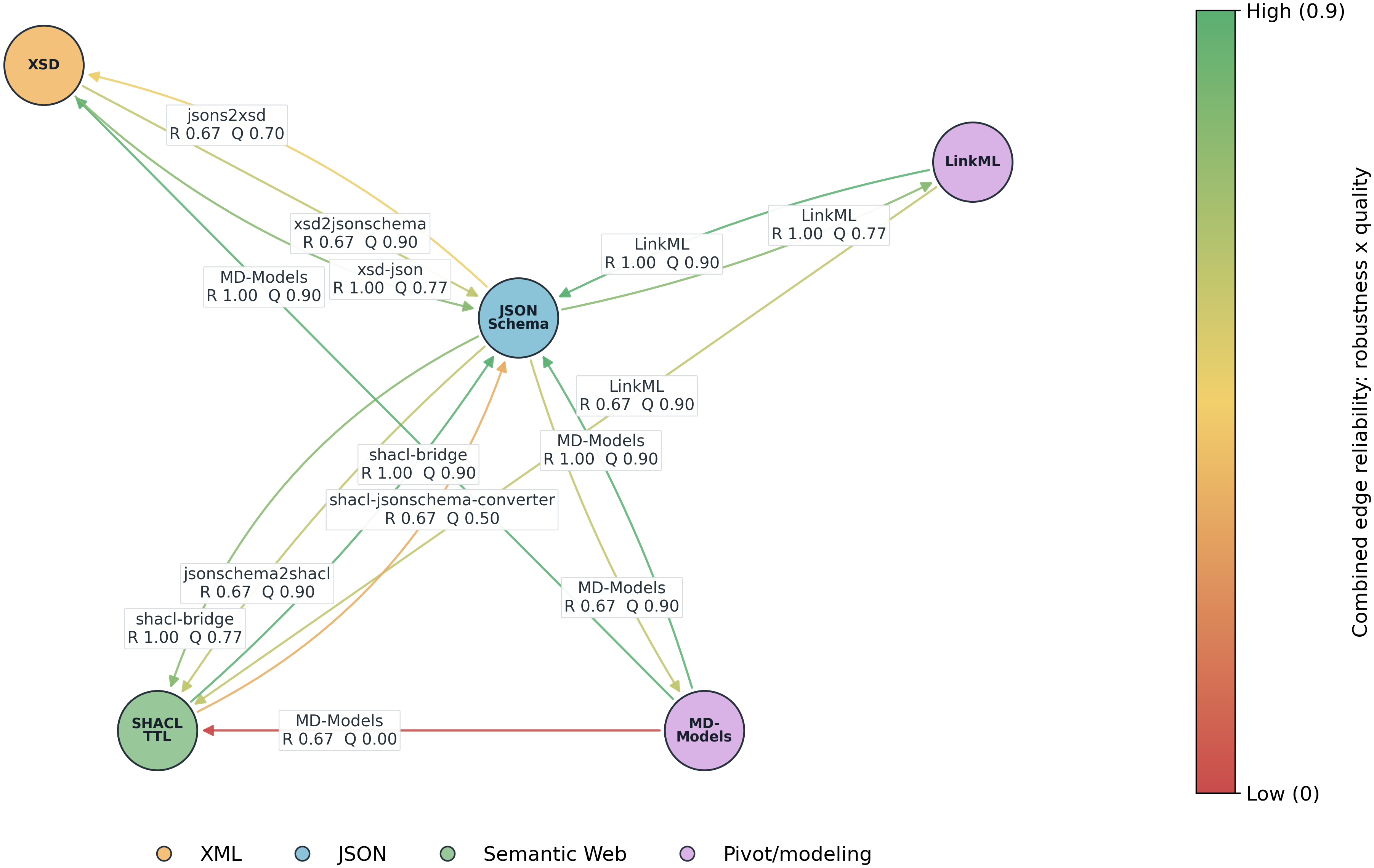}
\caption{Per-edge robustness and quality over the evaluated subset of the conversion graph. Edge labels show converter libraries together with robustness $R$ (automatic target-language validity rate) and conditional quality $Q$ (human-reviewed quality over automatically valid direct outputs). Edge color encodes the combined empirical edge reliability $R \cdot Q$ on the same $0$--$0.9$ red--yellow--green scale as the result matrix (Figure~\ref{fig:orchestrator_result_matrix}).
Across the direct one-step rows used here we annotated $42$ outputs over $14$ converter edges ($27$ G, $6$ L, $9$ I).
Unlike that matrix, this evaluates individual direct converter edges rather than best complete paths. The edge color combines both components as a reliability diagnostic, whereas the path ranking isejlf multiplies the conditional quality $Q$ along a path (Section~\ref{sec:edge_robustness}).}\label{fig:conversion_graph_edge_robustness}
\end{figure*}

The outputs are annotated with three labels: \textit{good} (G), meaning the result is valid and practically usable overall, even if minor details are imperfect; \textit{lacking} (L), meaning the result is valid and partially useful but misses important structure, constraints, or naming quality; and \textit{invalid} (I), meaning the path failed or produced an unusable result.
To reduce human bias and make the annotation reproducible, the first annotation pass is performed by a coding agent (Claude Code running Claude Sonnet~4.6) that inspects each input--output pair and assigns a label following written instructions published with the evaluation~\cite{darus_orchestrator_eval}.
Every agent label was then reviewed by a human using an interactive annotation viewer developed for this purpose, a browser-based tool displaying input and output side by side with a label and notes field.
Disagreements were resolved over several iterations, in most cases by correcting the individual label during review and, where a disagreement revealed an unclear criterion, by refining the instructions and re-running the agent on the affected outputs.
For example, the agent initially judged a structurally complete conversion as lacking only because the converter emits a bloated schema name (embedding the library name and a date); the instructions now direct it to judge the actual schema content and treat such naming artifacts as minor.
We use these labels at two levels: the result matrix (Figure~\ref{fig:orchestrator_result_matrix}) aggregates the best-ranked final outputs, while the edge-level plot (Figure~\ref{fig:conversion_graph_edge_robustness}) evaluates direct converter edges separately.
For the latter, we use only one-step rows in the final-output table: if a path consists of a single converter, the final output is also that converter's direct output on a real source-language schema.
Across the broad evaluation this yields $42$ direct outputs over $14$ evaluated converter edges ($27$ G, $6$ L, $9$ I).
Of these, $35$ produced a result and enter the conditional quality calculation; the seven failed conversions contribute to robustness but not to conditional quality.
Intermediate step outputs are still recorded for diagnostics and provenance, but they are not used for the default paper-facing edge metrics.

Figure~\ref{fig:orchestrator_result_matrix} summarizes the orchestrator-level result as a user would see it: only the best-ranked result per input contributes to a cell's quality, while the path count remains visible as context about graph connectivity.

Across the $60$ source--target--input tasks, the best-ranked result is good in $31$ cases, lacking in $12$, and invalid in $17$; that is, a usable (good or lacking) best result is surfaced for $43$ of the $60$ tasks.
Overall, the orchestrator executed $216$ final path attempts, of which $119$ produced a result and $97$ failed or produced no output.
These failures are not hidden: failed paths remain part of the returned diagnostics, including the failing converter step and error message, while successful alternatives are ranked and presented to the caller.
The empirical edge-quality ranking is decisive for this result: for pairs such as MD-Models~$\rightarrow$~JSON~Schema and XSD~$\rightarrow$~JSON~Schema, the best-ranked result is a usable direct conversion because the ranking favors the empirically higher-quality converter rather than whichever produces the largest output (Section~\ref{sec:edge_robustness}).
The strongest coverage is obtained for conversions involving JSON Schema, SHACL, LinkML, and MD-Models over direct or near-direct paths; the remaining invalid cells concentrate on conversions \emph{into} XSD or MD-Models from graph- or semantic-web sources (e.g., LinkML or SHACL), where the integrated converters provide no reliable path.

One instructive case is MD-Models~$\rightarrow$~SHACL.
Although MD-Models advertises a direct SHACL template, the evaluated direct edge is not a reliable structural schema conversion.
Inspection of the upstream MD-Models exporter\footnote{\url{https://github.com/FAIRChemistry/md-models/blob/main/src/exporters.rs}} shows that SHACL and ShEx generation first filters the model to objects with explicit semantic \texttt{Term} annotations; models without such annotations therefore produce an empty shapes graph.
This explains why the direct MD-Models~$\rightarrow$~SHACL edge is annotated as invalid in the review data, while the indirect MD-Models~$\rightarrow$~JSON~Schema~$\rightarrow$~SHACL path can still produce useful shapes: the JSON Schema template is driven by the structural MD-Models object model, and the subsequent SHACL converter receives a non-empty structural schema.

Running the same models through several target ecosystems also surfaced data-quality issues in otherwise valid-looking source schemas (for example, a schema name that is not a valid LinkML identifier and an undeclared Turtle prefix); the concrete cases are documented with the published evaluation data~\cite{darus_orchestrator_eval}.
We reported these cases to the respective upstream maintainers through issues and pull requests, and they have since been fixed in the source schemas, which is a concrete side benefit of exercising a schema across several ecosystems at once.
The roughly one quarter of tasks that still yield no usable best result reflect genuine gaps in the integrated converters, for which no path of sufficient quality exists; the orchestration and ranking themselves behave as intended. We discuss the implications in Section~\ref{sec:discussion}.

Figure~\ref{fig:conversion_graph_edge_robustness} visualizes the per-edge scores used by the empirical edge-quality ranking.

\subsection{Runtime Measurement}\label{sec:runtime}

To measure the runtime effect of sub-path caching, we run both registered SHACL~$\leftrightarrow$~JSON~Schema accuracy-benchmark directions once with caching enabled and once disabled while counting the complete evaluation runtime; the timing script and its results are part of the published evaluation data~\cite{darus_orchestrator_eval}, and the script can be rerun whenever converters or benchmarks change. The measurements were taken on a consumer laptop (Apple MacBook~Air with an M1 chip and 8\,GB of RAM).
For SHACL~$\rightarrow$~JSON~Schema, running all $39$ benchmark cases over all paths takes $55.5$\,s with caching and $62.6$\,s without, a reduction of $11.35\%$, because several attempted paths share the same intermediate SHACL JSON-LD serialization.
For JSON~Schema~$\rightarrow$~SHACL, the $88$ cases take $54.2$\,s with caching and $53.8$\,s without, a slowdown of $0.87\%$; there is still some overlap, but the benchmark is dominated by short paths and the cache bookkeeping slightly outweighs the reuse in this case.

\section{Integration into MetaConfigurator}\label{sec:mcintegration}

MetaConfigurator\footnote{\url{https://metaconfigurator.org}} is an open-source, JSON Schema-based web application for creating, editing, validating, and visualizing research data models, with form and code generation, schema inference, and AI-assisted mapping~\cite{neubauer2024metaconfigurator,neubauer2025datamodels,neubauer2025aiassisted}.

To make the orchestrator usable without any command-line interaction, we extend MetaConfigurator with schema-conversion import and export workflows.
The integration exposes the orchestrator through a user-friendly graphical interface with two entry points: an \emph{import} dialog, which converts a schema authored in another language (e.g., SHACL or XSD) into JSON Schema for further editing and visualization, and an \emph{export} dialog, which converts the current JSON Schema to a chosen target language. In both cases MetaConfigurator calls the orchestrator's REST API and presents the ranked list of resulting attempts.
By design, this integration only surfaces conversions that have JSON Schema, MetaConfigurator's native format, as either the source or the target; conversions directly between two other schema languages (e.g., XSD to SHACL) remain available through the orchestrator's API.

\begin{figure}[!t]
\centering
\includegraphics[width=1.0\columnwidth]{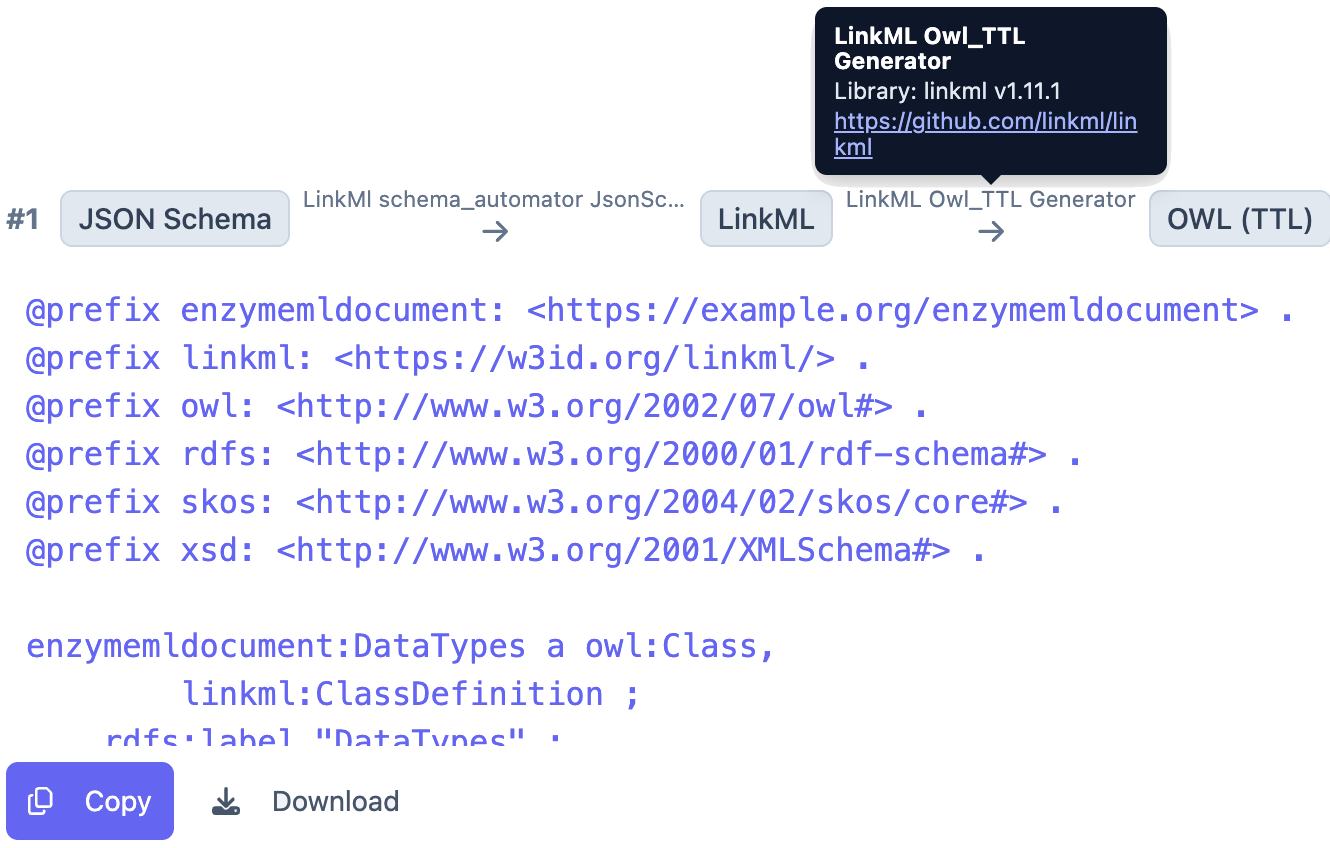}
\caption{Running and inspecting a conversion in MetaConfigurator: the export workflow shows a conversion result. Hovering over a path edge opens a tooltip with the converter library name, URL, and version.}\label{fig:mc-import}
\end{figure}

For each attempt, the interface renders the full conversion path as a sequence of language nodes connected by labeled edges, one edge per converter step (Figure~\ref{fig:mc-import}).
Because the orchestrator returns, for every step, the underlying converter library, URL, and version, hovering over an edge reveals this provenance together with a link to the corresponding library, so users can see precisely which tool produced a given result and can cite or report against it.
When a path fails, the interface highlights the responsible edge and shows the corresponding error message plus intermediate conversion results if applicable, giving users actionable diagnostics.
Successful attempts are shown with the converted schema and can be applied directly, with the best-ranked path presented first.

\section{Discussion}\label{sec:discussion}

\subsection{Implications of the results} The evaluation answers RQ1 constructively: orchestration can turn a fragmented set of imperfect converters into usable, reproducible conversion support for many practical cases.
Across the 60 evaluated tasks, the orchestrator surfaced a usable best result in 43 cases, including 31 directly usable conversions.
Its main contribution is therefore not that it improves individual converters, but that it changes the engineering workflow around them: users no longer have to find tools, assemble indirect chains, and compare failures manually.
They receive ranked alternatives, failed paths remain visible as diagnostics, and every step carries library, URL, and version provenance.
Even partially correct (``lacking'') outputs are useful in this setting, because they provide starting sketches that reduce the manual effort of porting a model.

The same results answer RQ2 by localizing the remaining limits.
The invalid cases are concentrated where no reliable path exists, especially conversions into XSD or MD-Models from graph- and semantic-web sources.
This is a property of the current converter landscape, not a failure of path orchestration.
For tool builders, the conversion graph and edge-level robustness measurements make these gaps actionable: they show where a new or improved converter would have the largest effect.
For practitioners, the coverage matrix makes current expectations explicit.
For both audiences, provenance and the published evaluation package~\cite{darus_orchestrator_eval} make conversions and converter comparisons more findable, reproducible, and reusable, aligning with the FAIR principles for data and research software~\cite{wilkinson2016fair,barker2022fair4rs}.
Because adding a language or converter requires only glue code and registration, the same graph also supports sustainable evolution: narrow or unmaintained converters can be diagnosed through provenance and failure reports, replaced, or supplemented by targeted new converters such as \emph{shacl-bridge}~\cite{shaclbridge,govindan2026shaclbridge}.

\subsection{Relation to model-driven engineering and pivot languages} The conversion graph is a lightweight, untyped counterpart to megamodels and transformation-chain frameworks in model-driven engineering, where systems and transformations are themselves modeled as graph elements~\cite{favre2005megamodel,wires,wimmer2012fact}.
Those frameworks assume a controlled technological space with a shared metametamodel; within that setting, prior work has studied chaining across incompatible metamodels and multi-objective selection of transformation chains~\cite{basciani2014chaining,momot}.
Our schema languages do not share such a foundation, so the orchestrator treats converters as black boxes and ranks paths by empirical quality rather than by static metamodel coverage.
This sacrifices analyzability of the individual transformation rules, but it allows reuse of ad-hoc tools embedded in different ecosystems.
Pivot frameworks such as LinkML~\cite{moxon2025linkml} and MD-Models~\cite{range2026mdmodels} remain complementary: they are attractive when users can adopt a single source language, while the orchestrator also serves cases where models already exist in several languages or where pivot generators are only one set of edges in a larger graph.

\subsection{Transferability and AI-assisted repair} The results also give a practical lesson for authors of portable schemas: the more a model relies on constructs specific to one schema language, the more loss to expect when translating it.
When portability is a design goal, authors should either stay within a transferable subset or use a pivot language explicitly designed for generation into several targets.
The boundary into semantic-web languages is harder still, because useful SHACL or OWL usually requires ontology terms that local schemas do not contain.
The MD-Models~$\rightarrow$~SHACL case illustrates this distinction: without explicit semantic \texttt{Term} annotations, the structural model remains present but the SHACL exporter emits no shapes.

Round trips could make this transferability measurable.
By converting a schema through paths that return to its source language and diffing the result against the original, the orchestrator could identify which constructs survive translation and which are consistently dropped.
The same observations delimit useful roles for LLMs.
They can act as fallbacks when no deterministic path yields a usable result, repair ``lacking'' outputs by filling the missing structure rather than performing the whole conversion, or infer ontology terms before a deterministic semantic-web exporter runs.
In this framing, LLMs complement reproducible converter chains instead of replacing them.

\subsection{Measuring and ranking conversion quality} Ranking paths requires a quality signal, but conversion quality is difficult to define.
Ground-truth target schemas are useful, yet constructing them requires conventions for choices that have no canonical answer, such as how XSD attributes or IRI-valued SHACL property paths should be named in JSON Schema.
Structural comparison is implementable and sufficient for ranking, but it penalizes semantically equivalent outputs with different structure.
Exact semantic equivalence can be decided for JSON Schema using witness generation~\cite{attouche2022witness}, but equivalence is binary while converter outputs are often partially correct.
Graded semantic similarity remains open; existing distances such as JSON edit distance~\cite{hutter2022jedi} are still structural, and language-specific comparison functions are needed for each target language.

The implemented ranking criteria should therefore be read as pragmatic baselines, not as definitive measures of semantic correctness or ecosystem quality.
The benchmark-backed scores demonstrate the mechanism where task-specific ground truth exists, and the empirical edge-quality estimate covers the rest of the graph.
Future work should add larger language-pair benchmarks, reviewed intermediate outputs for multi-hop chains, weighted or multi-objective scores~\cite{momot}, round-trip tests, and feature-sensitive ranking that scores only the constructs present in the current input.
As the graph grows, learned path selection could avoid executing historically weak paths.

\subsection{Threats to validity} Several limitations qualify these results.
The broad conversion matrix is primarily an integration and usability evaluation: it uses three inputs per source language and agent-assisted, human-reviewed G/L/I labels rather than task-specific ground truth, so it does not establish semantic correctness for every supported conversion.
Semantic accuracy is measured in depth only for SHACL~$\leftrightarrow$~JSON~Schema, and that benchmark uses minimal, isolated, hand-constructed cases with an equally weighted structural metric.
Because the benchmark was developed together with \emph{shacl-bridge}, its high score for that converter should not be interpreted as an independent comparison; complex nested schemas would be needed for that.
The complexity tiers and source line-count proxy are coarse, and the evaluation covers only single-file schemas.
Finally, weak graph regions such as OWL as a source and XSD as a target reflect the integrated converters available today.
These threats could be reduced by adding ground truth for more language pairs, systematically covering constraint combinations and nesting, refining metrics toward input-dependent accuracy, and evaluating multi-file schemas.

\section{Conclusion}\label{sec:conclusion}
We studied, empirically, to what extent existing, imperfect converters can be orchestrated into automated, reproducible, and quality-ranked conversions between heterogeneous schema languages (RQ1), and where the current converter landscape reaches its limits (RQ2).
Answering RQ1, the Schema Conversion Orchestrator shows that the orchestration is feasible and effective: schema languages become nodes, black-box converters become directed edges, and each request returns ranked conversion alternatives together with failure diagnostics and per-step library, URL, and version provenance.
Across the broad evaluation over five schema languages, it surfaces a usable best result for $43$ of the $60$ source--target--input tasks ($31$ of them directly usable), so orchestration delivers automated, reproducible, and quality-ranked conversions for roughly three quarters of the evaluated tasks.
Answering RQ2, the remaining quarter localizes the limits: the orchestration and ranking behave as intended there as well, and what is missing is a converter path of sufficient quality, with the gaps concentrated on conversions into XSD or MD-Models from graph- and semantic-web sources. The evaluation thus makes these gaps in the converter landscape explicit instead of hiding them.
The orchestrator's value lies in systematically finding, ranking, and transparently reporting the good conversions among many mediocre or failing ones, and in making the resulting conversions findable, reproducible, and easy to benchmark.
Because new languages and converters can be registered with little effort, the approach grows with the ecosystem it serves, and deterministic converter chains and LLM-based suggestions become complementary instead of competing.
Future work includes feature-sensitive and multi-objective ranking, learned path selection as the graph grows, task-specific ground truth for more language pairs, round-trip-based transferability analysis of schemas, support for multi-file and instance-level (data) transformations, and complementing deterministic converters with LLM-based suggestions for constructs that have no direct mapping.

\section*{Acknowledgments}
Financial support by the Deutsche Forschungsgemeinschaft (DFG, German Research Foundation) under grant numbers 390740016 (EXC 2075), and 441958208 (NFDI4Chem) is gratefully acknowledged.
During the preparation of this work the authors used Claude Code and ChatGPT Codex in order to support code development and to assist with language editing of the text. After using these tools/services, the authors reviewed and revised the content as needed and take full responsibility for the content of the published article.

\section*{Data Availability}
All evaluation inputs, scripts, and results are archived on DaRUS~\cite{darus_orchestrator_eval}, together with snapshots of the used versions of MetaConfigurator and the Schema Conversion Orchestrator.
The Schema Conversion Orchestrator and the MetaConfigurator integration of it are openly available on GitHub.\footnote{Schema Conversion Orchestrator: \url{https://github.com/MetaConfigurator/schema-conversion-orchestrator}; MetaConfigurator: \url{https://github.com/MetaConfigurator/meta-configurator}}

\bibliographystyle{preprinttran}
\bibliography{preprint-bibliography}

\end{document}